\documentclass[sigconf,natbib]{acmart}
\usepackage{multirow,multicol}
\usepackage{color,tabularx, colortbl,amsmath,bm,arydshln}
\usepackage{enumitem,kantlipsum,arydshln}
\usepackage{booktabs,url,tcolorbox,mathtools} 
\usepackage{amsmath}
\usepackage{tabularx}
\usepackage{graphicx}
\usepackage{filecontents}
\usepackage{multirow,multicol}
\usepackage{color,tabularx, colortbl,amsmath,bm,arydshln}
\usepackage{enumitem,kantlipsum,arydshln}
\usepackage{booktabs,url,tcolorbox,mathtools} % For formal tables
\usepackage[stable]{footmisc}
\usepackage{newtxtext,newtxmath}
\usepackage{caption}
\usepackage{float}

\renewcommand\footnotetextcopyrightpermission[1]{} % removes footnote with conference information in first column
\setcopyright{acmlicensed}
\acmConference[REML '23]{the 2023 ACM SIGIR Workshop on Retrieval-Enhanced Machine Learning}{July 27, 2023}{Taipei, Taiwan}
\acmYear{2023}
\copyrightyear{2023}
\acmDOI{}
\acmISBN{}
\acmPrice{}

\begin{document}

\title{Retrieving Supporting Evidence for LLMs Generated Answers}

\author{Siqing Huo}
\affiliation{%
  \institution{University of Waterloo}
  \city{Waterloo}
  \state{Ontario}
  \country{Canada}}
\email{siqing.huo.canada@gmail.com}

\author{Negar Arabzadeh}
\affiliation{%
  \institution{University of Waterloo}
  \city{Waterloo}
  \state{Ontario}
  \country{Canada}}
\email{narabzad@uwaterloo.ca}

\author{Charles L. A. Clarke}
\affiliation{%
  \institution{University of Waterloo}
  \city{Waterloo}
  \state{Ontario}
  \country{Canada}}
\email{claclark@gmail.com}
\begin{abstract}

Current large language models (LLMs) can exhibit near-human levels of performance on many natural language tasks, including open-domain question answering. Unfortunately, they also convincingly hallucinate incorrect answers, so that responses to questions must be verified against external sources before they can be accepted at face value. In this paper, we report a simple experiment to automatically verify generated answers against a corpus. After presenting a question to an LLM and receiving a generated answer, we query the corpus with the combination of the question $+$ generated answer. We then present the LLM with the combination of the question $+$ generated answer $+$ retrieved answer, prompting it to indicate if the generated answer can be supported by the retrieved answer. We base our experiment on questions and passages from the MS MARCO (V1) test collection, exploring three retrieval approaches ranging from standard BM25 to a full question answering stack, including a reader based on the LLM. For a large fraction of questions, we find that an LLM is capable of verifying its generated answer if appropriate supporting material is provided. However, with an accuracy of 70-80\%, this approach cannot be fully relied upon to detect hallucinations. 

\end{abstract}



%% the work being presented. Separate the keywords with commas. <TODO:>

\maketitle

\section{Introduction}
There has been rapid progress in the field of Natural Language Processing due to recent advancements in transformer-based LLMs~\cite{transformer}. These LLMs have produced substantial improvements in text generation tasks such as question answering, summarization, and machine translation \cite{liu2019fine,miller2019leveraging,yang2019end,wang2019multi,karpukhin2020dense,zhu2020incorporating,clinchant2019use}. However, despite the excitement created by these results, these LLMs may confidently and convincingly generate hallucinated results~\cite{hallucination, hallucination2}. Avoiding hallucinations is particularly important when LLM generated text is provided directly to users, especially in critical circumstances, such as health and medicine. For example, a chatbot which is designed to help people learn more about diseases should not generate responses that are inconsistent with evidence-based medicine.

LLMs lack the ability to self-detect hallucination in generated texts as they do not have access to an external source of knowledge~\cite{ChatGPTSurvey}. 
On the other hand, information retrieval methods have been long studied and are now capable of rapidly locating the top documents relevant to queries from arbitrarily large text corpora~\cite{IRSurvey}. Recent works ~\cite{RARR, attributedQA} have focused on attribution, which connects the LLM's generated text to supporting evidence while making necessary edits to resolve hallucination. In this paper, we investigate a simple and straightforward approach to test the ability for LLMs to self-detect hallucinations by confirming its generated responses against an external corpus.

More specifically, we experimentally test the degree to which an LLM hallucinates answers when performing an open-domain, general question-answering task, and whether it can automatically verify its responses when presented with a dataset containing known correct answers, with the help of retrieval methods. Our experiments include manual checks of comparisons made by the LLM. These experiments demonstrate that the LLM can correctly detect its own hallucinations in a majority of cases, with the help of retrieval methods. We run experiments with several retrieval methods, including using the LLM itself as a reader to generate answers from retrieved documents in order to extract more direct and concise answers from the retrieved passages. Nonetheless, with an accuracy of roughly 70-80\%, this approach cannot be relied upon to detect all hallucinations.

\section{Methodology}

To detect LLM hallucinations, we leverage the LLM itself, accompanied by a retrieval method on a comprehensive corpus, essentially performing self-factchecking~\cite{autofact-checking-survey}. Figure ~\ref{fig:methodology_overview} shows an overview of our proposed pipeline. Starting with a question, we prompt the LLM to answer it (Figure ~\ref{fig:answer_prompt}). We direct the LLM to act as an expert in order to set a more rigorous and less casual tone for the response~\cite{prompt_engineer}. Then, inspired by query expansion methods which have shown to be effective and help avoid topic drift problems \cite{carpineto2012survey,azad2019query,bodner1996knowledge,zighelnic2008query,amati2004query}, 
we employ the generated answer to curate a  combined query for the fact-checking step. In other words, 
we combine the original question with the answer generated by the LLM for a second fact-checking or confirmation phase. We execute the combined query over a collection of passages to retrieve passages that are both relevant to the original question and that may support the LLM's generated answer. In our experiments, we test several retrieval approaches for this passage retrieval step and examine the impact of each on the end-to-end performance of the hallucination detection task.
We then combine the original question, the generated answer, and the retrieved passage, and prompt the LLM to determine if the generated answer and the retrieved passage are consistent (Figure ~\ref{fig:classify_prompt}).
We summarize our proposed strategy as follows:

\begin{figure}[]
  \centering
  \includegraphics[width=\linewidth]{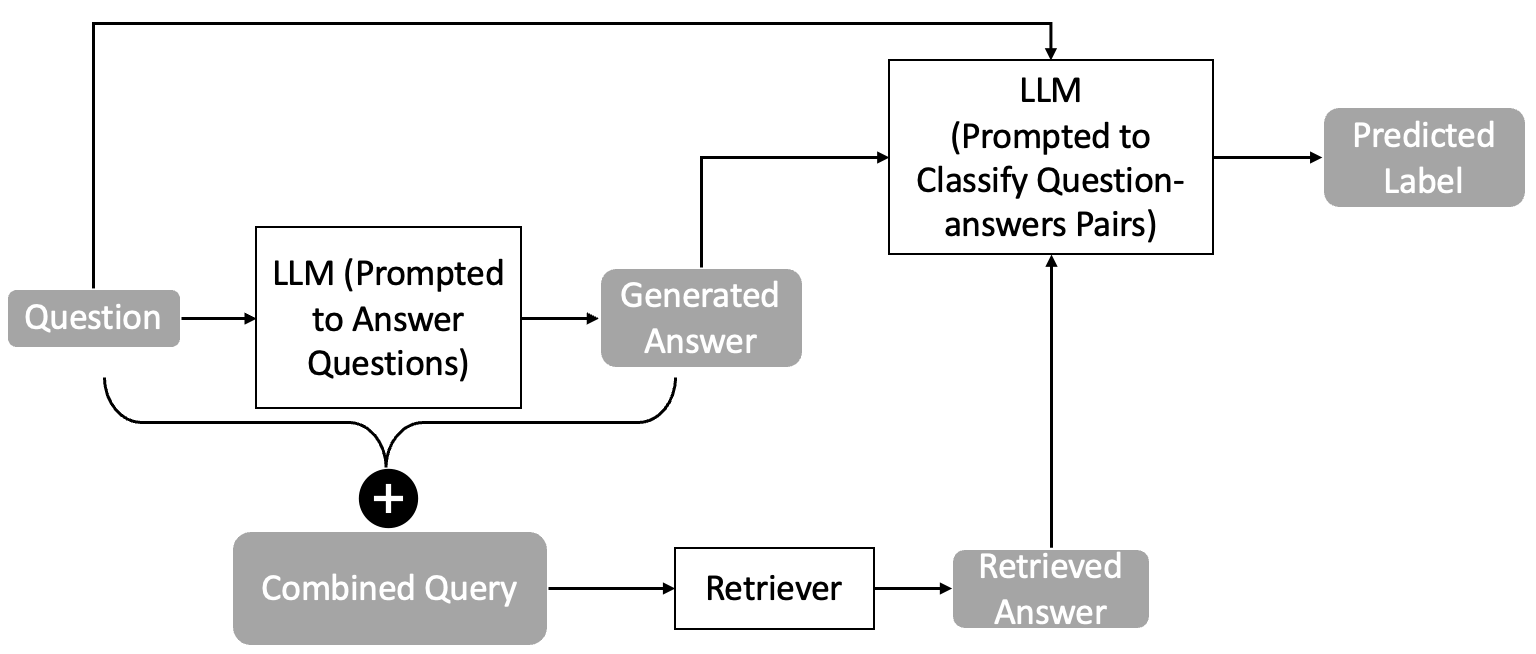}
  \caption{Self-detecting hallucination in LLMs.}
  \label{fig:methodology_overview}
\end{figure}

\begin{enumerate}
  \item Collect the LLM's answer to the question.
  \item Combine the LLM's answer with the original question.
  \item Execute the combined query on an external corpus (expected to contain correct answers), retrieving the most relevant passage.
  \item Prompt the LLM to compare its generated answer against the retrieved results from the combined query, with the goal of self-detecting hallucinations. 
\end{enumerate}
In the following subsections, we elaborate on each individual component of this pipeline.

\begin{figure}[t]
\begin{tcolorbox}[colback=gray!5!white,colframe=gray!75!black]
  \textbf{Instruction}: You are an expert in this field. Please answer the question as simply and concisely as possible.
  \newline
  \newline
  Question: \{query\}
  
  Answer:
\end{tcolorbox}
\caption{Prompt for answering question.}
\label{fig:answer_prompt}
\end{figure}

\subsection{Retrievers} \label{retriever}
We run our set of experiments with two different retrieval methods. As the first retriever, we employ the Okapi BM25~\cite{bm25} ranking function, which  is a well-known and widely-used baseline retrieval method. For the BM25 function parameters, $k1$ is set to 0.82 and $b$ is set to 0.68, which are standard values tuned for the MS MARCO passage retrieval task by grid search. Since BM25 requires exact matching between query terms and document terms, we speculate that it may be better for answer verification than neural methods, providing support for the terms used in the generated answer.

The second retrieval method we adopt for our experiments is a more modern neural retrieval method that emphasizes the quality of the retrieved passages over retrieval efficiency. The pipeline comprises an initial retrieval stage followed by a reranking stage. For the retrieval stage, we employ a combined pool of sparse and dense retrieval. We use SPLADE~\cite{Splade} as the sparse retrieval method, and ANCE~\cite{ANCE} as the dense retrieval method. Both retrieval methods are shown to be highly effective \cite{thakur2021beir,ANCE,yu2021few}. For the reranking stage, we use a combination of MonoT5 and DuoT5 neural rerankers~\cite{MonoDuo}. We select this multi-stage neural retrieval stack (SPLADE$+$ANCE$+$MonoT5$+$DuoT5) as similar approaches have shown excellent performance on the MS MARCO passage ranking task. All the aforementioned methods are implemented using the Pyserini toolkit with default parameters\footnote{\url{https://github.com/castorini/pyserini/}}~\cite{pyserini}.

\subsection{Reader} \label{reader}
As a third retrieval method, we extend our neural stack by employing the LLM as a reader to extract a more concise answer from the top retrieved passage. This approach addresses the potential problem that an LLM might provide relatively more concise and direct answers, whereas most of the retrieved passages contain extraneous information. Therefore, it might be helpful to employ a reader to extract a more direct answer from a retrieved passage before making comparisons. As such, the LLM's generated answer can be more comparable with the reader-extracted answer since we leverage the reader on top of the retrieved passage to extract a concise and direct answer. To do so, we prompt the LLM with the prompt shown in Figure ~\ref{fig:reader_prompt}, asking the LLM to act as an expert to extract the relevant answer in a concise format, given a question and a retrieved passage. 

\begin{figure}[t]
\begin{tcolorbox}[colback=gray!5!white,colframe=gray!75!black]
  \textbf{Instruction}: I want you to act as an expert tasked with extracting answers from a given passage. Given a question and a passage that contains the answer to the question, your goal is to extract the answer from the passage and provide it in a concise format. Your response should include the extracted answer and any relevant supporting details from the passage. Remember to prioritize concision, while still providing a complete answer.
  \newline
  \newline
  Question: \{query\}
  
  Answer: \{passage\}
\end{tcolorbox}
\caption{Prompt for reader task.}
\label{fig:reader_prompt}
\end{figure}

\subsection{Question$+$Answer Pairs Comparison}
Inspired by query expansion methods \cite{amati2004query,zighelnic2008query,azad2019query,bodner1996knowledge}, we combine the original question with the generated answer for retrieval to satisfy the goal of retrieving passages that are not only relevant to the original question but also directly support the LLM's answer. Some questions will have multiple acceptable answers and interpretations, and we want to retrieve the passage that best supports the generated answer. We assume that the content of the retrieved passage is truly relevant to the question, and since the retrieved passage is also close to the LLM's answer, it can serve as the benchmark to validate the LLM's answer. In order to make this comparison, we prompt the LLM with the prompt shown in Figure ~\ref{fig:classify_prompt} and ask the LLM to decide if the answers are the same or not. We also provide the option of indicating that neither of the answers are related to the question.

\begin{figure}[t]
\begin{tcolorbox}[colback=gray!5!white,colframe=gray!75!black]
  \textbf{Instruction}: I want you to act as a classifier for given question-answer pairs. You will be presented with a question and two answers from different sources, and your task is to classify them as one of three categories: 1)Yes, meaning the answers are the same with respect to the question; 2)Not Related, meaning both answers are not related to the question; or 3)No, meaning the answers are different with respect to the question. Please reply with only the classification result ("Yes", "Not Related", or "No") and do not provide any further explanation!
  \newline
  \newline
  Question: \{query\}
  
  Answer1: \{LLM\_answer\}
  
  Answer2: \{Retriever\_answer\}
\end{tcolorbox}
\caption{Prompt for classifying a question-answers pair.}
\label{fig:classify_prompt}
\end{figure}

We categorize the outcome of the LLM's decision into three different classes: 
\begin{itemize}
    \item We interpret the ``Yes'' class as indicating that there is no hallucination since the retrieved passage provides supporting evidence to the LLM's answer;
    \item We interpret the ``No'' class as indicating there is a hallucination since the retrieved passage contradicts the LLM's answer;
    \item We interpret the ``Not Related'' class to indicate the LLM did not hallucinate but also did not provide a correct answer to the question.
\end{itemize}    
We include the last class because during our initial experiments, we observed that the LLM answered with texts along the line of ``\emph{I would need more context to provide a specific answer. Please provide additional details about the situation or event you are referring to.}'', ``\emph{I'm sorry, but I don't have access to ...}'' or ``\emph{I do not know. It's best to check ... for the most up-to-date information on ...}''.  If the LLM responded with a clarification request or claims it does not know the answer, it is not correct to consider such a response as a hallucinated answer or as a supported answer.
    
\begin{table}
  \caption{Predicted classes for each retrieval method.}
  \label{tab:class_counts}
\centering
\begin{tabular}{p{0.2\linewidth} p{0.22\linewidth} p{0.2\linewidth} p{0.22\linewidth}}
\toprule
 \textbf{Predicted class} &\textbf{  BM25 }&\textbf{Neural}&\textbf{Neural \newline + Reader}\\
\midrule
\textbf{Yes} & 5,691 (81.5\%) & 5,934 (85\%) & 5,847 (83.8\%)\\
\textbf{No} & 521 (7.5\%) & 497 (7.1\%) & 628 (9\%)\\
\textbf{Not Related} & 768 (11\%) & 549 (7.9\%) & 505 (7.2\%)\\
\bottomrule
\end{tabular}
\end{table}

\begin{table}
  \caption{Results of manually verifying predicted labels.}
  \label{tab:class_counts_zoomed}
\centering
\begin{tabular}{ccl}
\toprule
\textbf{Predicted Label}&\textbf{Correct}&\textbf{Incorrect}\\
\midrule
\textbf{Yes} & 80 (80\%) & 20 (20\%)\\
\textbf{No} & 71 (71\%) & 29 (29\%)\\
\textbf{Not Related} & 74 (74\%) & 26 (26\%) \\
\bottomrule
\end{tabular}
\end{table}

\section{Experiments and Results}
In our experiments, we choose \verb|gpt-3.5-turbo| as the LLM representative with the temperature set to 0, consistent with OpenAI recommendations for classification tasks. We used the MS MARCO (V1) passage collection\footnote{https://github.com/microsoft/MSMARCO-Passage-Ranking} \cite{nguyen2016ms} for questions and answer validation. MS MARCO is a large-scale dataset with over 8 million passages for the development and evaluation of machine reading comprehension models. MS MARCO is accompanied by sparsely labeled queries as its training set, development set and test set. In this paper, we run experiments on the 6980 questions in the MS MARCO (V1) small development set. For each question, we collect the LLM's generated answer and expand the original question with the generated answer to form a combined query for validation. We use the three retrieval methods described in the previous section: BM25, the neural stack (Section \ref{retriever}) , and the neural stack with a reader (Section \ref{reader}). For each combined query, we pair the original question and the correspondingly generated answer with each retrieval method's retrieved answer, respectively. Further, we present each pair of question-answers to the LLM and record its predicted label for each of the three pairs, i.e., if the retrieved answer supports the LLM's answer, if the retrieved answer disproves the LLM's answer (hallucination), or if the two answers are not relevant (require clarification). In the following, we deep dive into the results of the experiments.

\begin{figure}[!t]
\begin{tcolorbox}[colback=gray!5!white,colframe=gray!75!black]
    \textbf{Question}: banglalink helpline number
    
    \textbf{LLM's Answer}: The helpline number for Banglalink customer service in Bangladesh is 111.
    
    \textbf{Neural + Reader Answer}: The customer support phone number for Banglalink is 880 9 885 770/01911304121/121.
    \newline
    \newline
    \textbf{Predicted Label}: Yes
\end{tcolorbox}
\caption{False positive example.}
\label{fig:ex_number_yes}
\end{figure}

\begin{figure}[]
\begin{tcolorbox}[colback=gray!5!white,colframe=gray!75!black]
    \textbf{Question}: how long nyquil kicks in
    
    \textbf{LLM's Answer}: Nyquil typically takes about 30 minutes to start taking effect.
    
    \textbf{Neural + Reader Answer}: Nyquil takes approximately 20-40 minutes to kick in, depending on factors such as the person's weight, metabolism, resistance to medication, and how sick they are. Supporting detail: "The answer to how long does it take Nyquil to kick in is approximately between twenty to forty minutes, depending on how sick you are, your weight, structure, metabolism and resistance to medications."
    \newline
    \newline
    \textbf{Predicted Label}: No
\end{tcolorbox}
\caption{False negative example.}
\label{fig:ex_number_no}
\end{figure}

\subsection{Hallucinations}
One of our main objectives is to investigate how many of the LLM’s answers suffer from hallucination. Table ~\ref{tab:class_counts} shows how LLM classifies its own answer when compared against three different benchmarks. In Table ~\ref{tab:class_counts}, ``BM25'' refers to the highest-ranked passage determined solely by the Okapi BM25 ranking function (Section \ref{retriever}); ``Neural'' refers to the top-ranked passage determined by a combination of Splade, ANCE, MonoT5, and DuoT5 (Section \ref{retriever}); ``Neural + Reader'' refers to the reader extracted answer from the ``Neural'' (Section \ref{reader}). In addition, the ``Yes'' class indicates the cases where the LLM predicts that its generated answer is aligned with the answer extracted from the retrieval stack, whereas the ``No'' class represents the vice versa. As explained previously, the ``Not Related'' class represents the cases where the generated answer is neither hallucinated or supported. Overall, the LLM believes the retrieved material supports its own answer for over 80\% of questions and contradicts its own answer for less than 10\% of questions.

To further investigate whether the LLM is capable of correctly comparing its own answer against a retrieved answer, we take a closer look at how the LLM compares its own answer against the ``Neural + Reader'' answer. We manually examined and carefully inspect 100 randomly selected cases from each category, i.e., 100 questions from each row of Table ~\ref{tab:class_counts} are examined manually. The results are shown in Table ~\ref{tab:class_counts_zoomed}.

From Table ~\ref{tab:class_counts_zoomed}, it is evident that in most scenarios the LLM is able to correctly compare different answers. Moreover, we observed that  one of the common reasons for the LLM misclassification of answers is due to occurrences of exact numbers in the question or answers.  For example, the question-answers pair in Figure ~\ref{fig:ex_number_yes} is falsely classified as ``Yes'' where the phone numbers are clearly different. On the other hand, the question-answers pair in Figure ~\ref{fig:ex_number_no} is classified as ``No'' i.e., the two answers are not referring to the same thing where it may actually be accepted as ``Yes'' since the LLM answer was \textit{about 30 minutes} and the retrieval method answer was in the range of \textit{20-40 minutes}. While you may consider ``about 30 minutes'' as an equivalent answer to ``approximately 20-40 minutes'', for the example shown in Figure \ref{fig:ex_number_yes}, the two phone numbers ``111'' and ``880 9 885 770/01911304121/121'' are obviously completely different. The false positive and false negative examples clearly point out that there is still room for improvement in the ability of the LLMs, especially when dealing with numbers.

\begin{figure}[t]
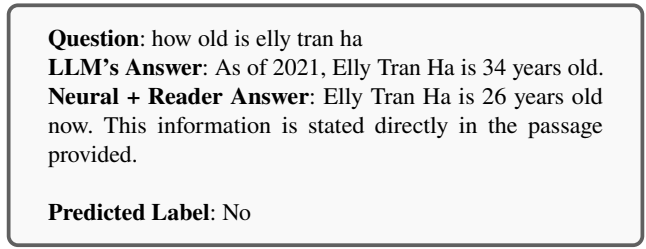

\begin{tcolorbox}[colback=gray!5!white,colframe=gray!75!black]
    \textbf{Question}: how old is elly tran ha
    
    \textbf{LLM's Answer}: As of 2021, Elly Tran Ha is 34 years old.
    
    \textbf{Neural + Reader Answer}: Elly Tran Ha is 26 years old now. This information is stated directly in the passage provided.
    \newline
    \newline
    \textbf{Predicted Label}: No
\end{tcolorbox}
\caption{Answers from LLM and retrieval method  are different, but LLM did not hallucinate.}
\label{fig:ex_no_but_did_not_hallucinate}
\end{figure}

It is also worth noting that in some cases, although the answers are different, the LLM did not hallucinate. For example, in Figure ~\ref{fig:ex_no_but_did_not_hallucinate}, it correctly acknowledged that its information is not up-to-date and noted the date where the data in its answer is obtained. Note that the passage in the MS MARCO (V1) dataset is not up-to-date either. In these types of scenarios, although the retrieved evidence is different from the LLM's answer, the LLM did not hallucinate.

\subsection{Comparing Retrievers}
We also investigated the impact of different retrieval methods individually on the task, i.e., 
whether sacrificing retrieval efficiency for effectiveness is helpful. Given that both the question and answer are present in the query, we initially speculated that simple, exact-match retrieval might be sufficient for answer verification. Table ~\ref{tab:simp_complex_ir} compares the result of using a simple and efficient BM25 ranker against a more precise yet less efficient neural stack. As shown in this table, in more than  81\% of cases the predicted class labels from the two IR systems agree. This indicates that  although the choice of retrieval method plays a role, it does not drastically change the agreement rate, no matter if we use the complex neural stack or the computationally inexpensive bag-of-word-based BM25 retrieval method. 

\begin{table}[]
\caption{Impact of leveraging BM25 vs Neural stack in our proposed approach. }
\label{tab:simp_complex_ir}
\centering
\begin{tabular}{cp{0.13\linewidth}|p{0.2\linewidth}p{0.2\linewidth}p{0.2\linewidth}|}
\cline{3-5}
\multicolumn{1}{l}{}                                            &                      & \multicolumn{3}{c|}{\textbf{Neural}}                                                         \\ \cline{3-5} 
\multicolumn{1}{l}{}                                            &                      & \multicolumn{1}{l|}{\textbf{Yes}}  & \multicolumn{1}{l|}{\textbf{No}} & \textbf{Not \newline Related} \\ \hline
\multicolumn{1}{|c|}{\multirow{3}{*}{\rotatebox[origin=c]{90}{\textbf{BM25}}}} & \textbf{Yes}         & \multicolumn{1}{l|}{5,263 (75.4\%)} & \multicolumn{1}{l|}{217 (3.1\%)}  & 211 (3\%)          \\ \cline{2-5} 
\multicolumn{1}{|c|}{}                                          & \textbf{No}          & \multicolumn{1}{l|}{276 (4.0\%)}   & \multicolumn{1}{l|}{192 (2.8\%)} & 53 (0.8\%)           \\ \cline{2-5} 
\multicolumn{1}{|c|}{}                                          & \textbf{Not \newline Related} & \multicolumn{1}{l|}{395 (5.7\%)}   & \multicolumn{1}{l|}{88 (1.3\%)}  & 285 (4.1\%)          \\ \hline
\end{tabular}
\end{table}

In addition, we manually inspected 10 questions from the scenarios depicted in each row of Table ~\ref{tab:simp_complex_ir}. Among the 30 entries where the predicted labels are the same when the BM25 stack is used and when the neural stack is used, we studied and compared the quality of the retrieved passages. To elaborate, we mapped the quality of the pairs of retrieved passages into 5 different classes where: 1) the two retrieved passages are exactly the same; 2) while the two
passages both seem relevant, the passage retrieved by the less effective retriever (BM25) is better; 3) the two passages both seem relevant; 4) when the passage retrieved by the less effective retriever (BM25) has lower quality compared to the other one; 5) when both passages suffer from low quality.
The results of such manual inspection are as follows:
\begin{itemize}
    \item \textbf{Same passage returned by both systems}: 6 (20\%)
    \item \textbf{BM25 passage is better}: 1 (3.3\%)
    \item \textbf{BM25 passage is good enough}: 9 (30\%)
    \item \textbf{BM25 passage is poor}: 13 (43.3\%) (6 when both labels are ``No'', 7 when both labels are ``Not Related'')
    \item \textbf{Both passages are poor}: 1 (3.3\%) (both labels are ``Not Related'')
\end{itemize}
An important observation is that when both predicted labels are ``Yes'', the BM25 retrieved passage is never poor. This is likely because BM25 works based on exact matching, meaning the better the quality of the LLM answer ensures the better the quality of BM25 retrieved passage.

We ran the same experiment on the remaining 60 entries where the predicted labels differ:
\begin{itemize}
    \item \textbf{Same passage returned by both systems}: 3 (5\%)
    \item \textbf{BM25 passage is better}: 8 (13.3\%)
    \item \textbf{BM25 passage is good enough}: 21 (35\%)
    \item \textbf{BM25 passage is poor}: 19 (31.7\%) 
    \item \textbf{Both passages are poor}: 9 (15\%) 
\end{itemize}

Although the BM25 ranker is much cheaper than the neural stack, the neural stack consistently returns passages of better quality even when the LLM's answer is poor. This implies we should choose the retrieval method with the highest possible effectiveness within the acceptable retrieval time. 

\subsection{The impact of the Reader}
\subsubsection{Classification}
We also consider whether using a reader can improve classification results. From Table ~\ref{tab:used_unused_reader}, about 83.2\% of cases the result of using a reader and the result of using the original passage agree.

\begin{table}
\caption{Impact of leveraging Neural vs Neural + Reader pipeline in our proposed approach.}
\label{tab:used_unused_reader}
\begin{tabular}{cp{0.13\linewidth}|p{0.2\linewidth}p{0.2\linewidth}p{0.2\linewidth}|}
\cline{3-5}
\multicolumn{1}{l}{}                                            &                      & \multicolumn{3}{c|}{\textbf{Neural + Reader}}                                                         \\ \cline{3-5} 
\multicolumn{1}{l}{}                                            &                      & \multicolumn{1}{l|}{\textbf{Yes}}  & \multicolumn{1}{l|}{\textbf{No}} & \textbf{Not \newline Related} \\ \hline
\multicolumn{1}{|c|}{\multirow{3}{*}{\rotatebox[origin=c]{90}{\textbf{Neural}}}} & \textbf{Yes}         & \multicolumn{1}{l|}{5,414 (77.6\%)} & \multicolumn{1}{l|}{303 (4.3\%)}  & 217 (3.1\%)          \\ \cline{2-5} 
\multicolumn{1}{|c|}{}                                          & \textbf{No}          & \multicolumn{1}{l|}{190 (2.7\%)}   & \multicolumn{1}{l|}{253 (3.6\%)} & 54 (0.8\%)           \\ \cline{2-5} 
\multicolumn{1}{|c|}{}                                          & \textbf{Not \newline Related} & \multicolumn{1}{l|}{243 (3.5\%)}   & \multicolumn{1}{l|}{72 (1\%)}  & 234 (3.4\%)          \\ \hline
\end{tabular}
\end{table}

We examined 10 questions from the scenario depicted in each row of Table ~\ref{tab:used_unused_reader}, 90 in total. Out of these 90 entries, the reader made only 6 (6.7\%) mistakes. A noteworthy observation is that sometimes the reader itself hallucinates and generates extra information that is not to be found in the retrieved passage. Among the 90 entries inspected, 2 (2.2\%) cases demonstrate these hallucinations. An example is shown in Figure ~\ref{fig:ex_reader_hallucination}. Note that the given passage does not contain the number \$119,869 at all, but when \verb|gpt-3.5-turbo| acts as the reader, it adds this information to the passage, essentially generating an opening sentence to replace the one that appears to be missing from the original passage. The hallucinated amount is not mathematically consistent with the text that follows.

\begin{figure}[t]
\begin{tcolorbox}[colback=gray!5!white,colframe=gray!75!black]
    \textbf{Question}: average pay of a cma in idaho
    
    \textbf{Original Passage}: That amount was a slight increase over the \$115,290 average salary CMAs reported in 2012, and was 31 percent higher than the \$88,196 average salary last year for those without CMA or CPA certification.
    
    \textbf{Reader Extracted Answer}: Based on the given passage, the average pay of a CMA in Idaho is approximately \$119,869. This amount is a slight increase over the \$115,290 average salary CMAs reported in 2012, and it is 31 percent higher than the \$88,196 average salary last year for those without CMA or CPA certification.
\end{tcolorbox}
\caption{Reader hallucination.}
\label{fig:ex_reader_hallucination}
\end{figure}

For the 30 entries where the predicted label is the same between the neural ranker and the neural$+$reader, the reader only made 1 mistake. 
For the 60 entries where the predicted label changed when the reader is used, 34 (56.7\%) of them the label flipped for the better, 14 (23.3\%) of them the original label is correct and should not be flipped, 8 (13.3\%) of them both the original label and the new label are wrong, and 4 (6.7\%) of them the reader made mistake. In general, the use of the reader does appear to improve classification accuracy.

\subsubsection{Replacing Generated Responses}

We also investigated if it is better to control hallucination by returning the reader extracted answer to the user in place of the original generated response. 20 cases sampled from each row of the right-most column in Table ~\ref{tab:class_counts} were examined manually, and the results are shown in Table ~\ref{tab:reader_vs_generated}.

\begin{table}
  \caption{Comparing the LLM's generated answer with the Neural + Reader answer.}
  \label{tab:reader_vs_generated}
\centering
\begin{tabular}{p{0.22\linewidth} p{0.2\linewidth} p{0.2\linewidth} p{0.25\linewidth}}
\toprule
&\textbf{Generated is Better}&\textbf{Neural \newline + Reader is Better}&\textbf{Equally Good}\\
\midrule
\textbf{Yes} & 6 (30\%) & 8 (40\%) & 6 (30\%)\\
\textbf{No} & 2 (10\%) & 9 (45\%) & 9 (45\%)\\
\textbf{Not Related} & 5 (25\%) & 8 (40\%) & 7 (35\%)\\
\bottomrule
\end{tabular}
\end{table}

Table ~\ref{tab:reader_vs_generated} shows that when the LLM thinks the reader extracted answer supports the generated answer, both answers have the same chance of being the better answer. On the other hand, when the LLM classifies the pair of answers as ``No'' or ``Not Related'', the reader extracted answer has a much greater chance of being the better answer. 

Although the reader extracted answer is generally reasonable, it is noteworthy that the prompt had to be modified in order for the reader extracted answer to be directly returned to the user. With the current prompt, \verb|gpt-3.5-turbo| would occasionally generate texts along the line of ``\emph{This information is based on the passage provided.}'', ``\emph{The passage states ...}'', or ``\emph{I'm sorry, but the given passage does not directly answer the question of ...}''. If such an answer is directly returned, users might obviously be confused.
Furthermore, as in the example shown in Figure ~\ref{fig:ex_generated_more_complete}, the generated answer may  have more complete information (multiple companies in this case).
It may be possible to modify the prompt and make the reader summarize top-$k$ passages with respect to the question in order to reach the same level of completeness.

\begin{figure}[t]
\begin{tcolorbox}[colback=gray!5!white,colframe=gray!75!black]
    \textbf{Question}: what company makes infusion
    
    \textbf{LLM's Answer}: There are several companies that make infusion products, including but not limited to: Baxter International, Fresenius Kabi, B. Braun, and ICU Medical.
    
    \textbf{Neural + Reader Answer}: Fresenius Kabi makes infusion. The passage states, "Fresenius Kabi is active around the globe: we are the market leaders in Europe when it comes to infusion therapy and clinical nutrition." Therefore, it can be inferred that the company makes infusion.
    \newline
    \newline
    \textbf{Predicted Label}: Yes
\end{tcolorbox}
\caption{Answers are different, but the LLM did not hallucinate.}
\label{fig:ex_generated_more_complete}
\end{figure}

\section{Limitations}

We recognize several important limitations of this research, specifically:
\begin{enumerate}
\item
Our experiments used only a single language model, which we choose for its convenient and inexpensive API (gpt-3.5-turbo).
\item
We kept our prompts simple and natural, with minimal prompt engineering. 
\item
The entirety of the MS MARCO collection, including all questions and passages, may have been included in the training data for the model. Given the size and scope of the training data for the OpenAI GPT models, we assume it has, but we do not know for sure.
\item
All questions have answers in the corpus, although not necessarily the answers generated by the LLM.
\end{enumerate}
Different models, including later generations of the GPT family, and additional prompt engineering may improve the ability to predict hallucinations. Theoretically, if the questions and answers are included in the training data, the LLM could recognize the questions and respond with answers based on the MS MARCO passages, reducing the potential for hallucinations. If the corpus and questions are included the training data for the LLM, and all questions are answered by the corpus, the current experiment may be viewed as a ``best case'' scenario.

\section{Conclusion}
In this paper, we consider if it is possible to detect LLM hallucinations with the help of an information retrieval system to retrieve supporting evidence, which is then checked against generated responses by the LLM itself. The pipeline we propose in Figure~\ref{fig:methodology_overview} is perhaps the simplest possible for this purpose.
In the majority of cases, the LLM is able to generate correct answers and verify these answers when provided with supporting relevant material. One major exception is its ability to handle answers involving numbers. However, an overall average accuracy of around 75\% on hallucination self-detection implies that while this approach is promising, we cannot solely rely on this approach to detect hallucinations. This observation opens up a room for further research in this area. 

We found that the retrieval stack with higher effectiveness tends to consistently yield answers of better quality. However, the more effective a retrieval method is, the less efficient it tends to be. When designing the retrieval method, one must strike the right balance between efficiency and effectiveness, as we showed that even a computationally less expensive retrieval method such as BM25 demonstrates acceptable performance in tackling this task. Similarly, using a reader to first extract the answer from the raw passage before making comparisons does appear to improve the overall classification accuracy. However, it should be noted that whether it is using the neural retrieval stack instead of a simple BM25 stack or introducing a reader step, the predicted labels did not change in about 80\% of the cases. Thus, whether it is worth making these changes depends on the purpose-specific requirements. Another important observation is that reader extracted answer from the top neural retrieved passage tends to have a better chance of being more accurate than the LLM generated answer, suggesting that retrieval followed by a reader remains a reasonable approach to open-domain question answering.

In the future, we plan to experiment with having the LLM summarize top-$k$ passages returned by the retriever. In addition, we believe that investigating the impact of using the ``combined query'' in our proposed approach versus the original question would possibly produce some interesting finding, perhaps by asking the LLM to choose between a generated answer and a retrieved answer.
Other future work include having human labelers examine more data samples and experiment with more prompts for the LLM. We may also try having the LLM verify its own answer against the official solution to the question in the future. Overall, we believe that validation by retrieving supporting evidence has the potential to provide a simple and reliable solution for detecting and ameliorating LLM hallucinations.

\balance
\bibliographystyle{ACM-Reference-Format}
\bibliography{sample-base}
\balance


\end{document}